\documentclass[11pt]{article}
\usepackage{jheppub} 

\usepackage{braket,slashed,bm}
\usepackage{array,multirow}
\usepackage[normalem]{ulem}
\usepackage{xcolor,cancel}
\usepackage{amsmath}
\usepackage{bbold}
\usepackage{dsfont}
\usepackage{cleveref}
\usepackage{xspace}
\usepackage{verbatim}
\usepackage{soul}

\crefname{table}{Table}{Tables}
\crefname{equation}{Eq.}{Eqs.}
\crefname{appendix}{App.}{Apps.}
\crefname{section}{Sec.}{Secs.}
\crefname{figure}{Fig.}{Figs.}

\usepackage{physics}
\usepackage{stackengine,xcolor,scalerel}
\usepackage{subcaption}

\usepackage[T1]{fontenc} 
\usepackage{mathrsfs}

\usepackage{tikz}
\usetikzlibrary{arrows,decorations.pathmorphing,backgrounds,positioning,fit,petri,automata,shadows,calendar,mindmap,
decorations.markings,calc}

\newcommand{\gev}{\text{GeV}}
\newcommand{\tev}{\text{TeV}}

\title{A $W^\pm$ polarization analyzer from Deep Neural Networks }

\author[a]{Taegyun Kim} 
\author[a]{and Adam Martin}
\affiliation[a]{Department of Physics, University of Notre Dame, South Bend, IN
  46556 USA} 

\abstract{ In this paper we train a Convolutional Neural Network to classify longitudinally and transversely polarized hadronic $W^\pm$ using the images of boosted $W^{\pm}$ jets as input. The images capture angular and energy information from the jet constituents that is faithful to properties of the original quark/anti-quark $W^{\pm}$ decay products without the need for invasive substructure cuts. We find that the difference between the polarizations is too subtle for the network to be used as an event-by-event tagger. However, given an ensemble of $W^{\pm}$ events with unknown polarization, the average network output from that ensemble can be used to extract the longitudinal fraction $f_L$. We test the network on Standard Model $pp \to W^{\pm}Z$ events and on $pp \to W^{\pm}Z$ in the presence of dimension-6 operators that perturb the polarization composition.
}

\begin{document}
\maketitle

\section{Introduction and Motivation}

We are entering the precision LHC era. No light new particles have been seen to date, and while it is not impossible that the full run of the LHC will expose a new particle, we must consider the possibility that new physics is simply too heavy to produce substantially at the LHC. In this scenario, the search for new physics moves from obvious and direct -- spectacular signals of on-shell particle production, such as resonant peaks or large missing energy signatures -- to indirect and subtle, looking for deviations in distributions from the Standard Model (SM) prediction.

The polarization of massive gauge bosons is an interesting avenue to explore using the indirect approach. The transverse and longitudinal fractions vary depending on what process (e.g. single boson production versus diboson) and energy are considered, and are a detailed probe of the machinery of the Standard Model (SM).  Moreover, the longitudinal polarizations of the $W^{\pm}/Z$ are especially sensitive of the mechanism of electroweak symmetry breaking, as perturbative unitarity in longitudinal boson scattering can be maintained only through a delicate balance of contributions~\cite{Lee:1977eg,Chanowitz:1985hj}. In scenarios where the Higgs properties deviate even slightly from the SM expectations, such as in composite Higgs scenarios~\cite{Dugan:1984hq, Agashe:2004rs, Giudice:2007fh}, this balance breaks down and we expect dramatic signals. 

If the scale of new physics is light, these signals usually take the form of resonances. However, if the scale of new physics is heavy, its effects can be captured by an effective Lagrangian, the SM augmented by a series of higher dimensional operators. The imprint of UV physics is left on the pattern of operators -- the relative size and type of operator generated. Within the effective Lagrangian language, gauge bosons can appear either as field strength $F_{\mu\nu}$ or in the covariant derivatives of Higgs fields $D_{\mu} H$. The former are transversely polarized while the latter are (primarily) longitudinally polarized.

To disentangle the effects from the two types of operators, we need to differentiate polarizations. The polarization difference is the clearest if one can reconstruct the $W^\pm/Z$ and boost back to its rest frame, therefore current polarization studies have been restricted to leptonic final states~\cite{Chatrchyan:2011ig, Aad:2012ky}. However, while leptonic final states are clean, they suffer from low branching ratios and ambiguities due to the presence of neutrinos.

The goal of this paper is to develop a polarization analyzer for hadronic $W^\pm$ using machine learning tools. In order to avoid huge backgrounds and combinatorial issues, we focused on analyzing the polarization of boosted  $W^\pm$. Boosted $W^\pm$ have collimated decays, so they look like a single fat ($\Delta R \sim 1$) jet at detector level. As all the decay products are (theoretically) contained within the fat jet, this mitigates the headache of reconstructing the $W^{\pm}$ and provides several useful handles, exploited through jet substructure techniques, at distinguishing the hadronic $W$ from a QCD jet~\cite{Thaler:2008ju, Kaplan:2008ie, Almeida:2008yp,  Butterworth:2008iy, Thaler:2010tr}. For our network input, we use images of the the fat $W$ jets (preprocessed and pixelized) rather than specific substructure variables. Then, using event samples where one polarization completely dominates for training, the resulting network is able to pick up on how  the polarization of the boosted $W$ is manifest in subtle image differences. For transverse $W^\pm$, we use $W+\text{jets}$ as the training sample, while for longitudinal $W^\pm$ we use a heavy Higgs $H \to W^+W^-$.

Machine learning techniques have previously been applied to hadronic $W$ in Ref.~\cite{Grossi:2020orx}, focusing on extracting the polarization in semi-leptonic $W$ produced in vector boson fusion, $pp \to W(\ell\nu)W(jj) + jj$ and showing promising results. Comparing our approach with theirs, we use jet images, while Ref.~\cite{Grossi:2020orx} used the four vectors of the lepton and jets as the input to the network. More importantly, the simulation in Ref~\cite{Grossi:2020orx} consisted of parton-level events smeared with detector efficiencies and without a genuine parton shower. As we will discuss in more detail below, the main difficulty with the hadronic $W^\pm$ (in general, and for a polarization study in general), comes from extra radiation, specifically in identifying the quarks the $W^\pm$ decays to and their accompanying radiation, and weeding out extraneous radiation. As smeared parton-level events will never generate extra radiation, it is hard to extrapolate the results of Ref.~\cite{Grossi:2020orx} to a realistic collider environment.

The layout of the rest of this paper is as follows. In Sect.~\ref{sec:partonlevel} we review the parton level observables, in the lab and $W$ rest frames, that are sensitive to polarization information. Next, in Sect.~\ref{sec:network}, we describe our neutral network structure, training samples, and performance. Our main results are contained in Sec.~\ref{sec:results}, broken up into two subsections: i.) results for SM $pp \to W^\pm Z$ production, Sec.~\ref{sec:SMWZ} and ii.) $pp \to W^\pm Z$ production in the presence of higher dimensional operators that alter the polarization fraction, Sec.~\ref{sec:SMEFTresults}. Section~\ref{sec:discussion} contains our conclusions.

\section{$W^\pm$ polarization at parton level}
\label{sec:partonlevel}

In the rest frame of a $W^\pm$ boson, the decay products from the longitudinal and transverse polarizations have different angular distributions. Taking the decay products to be massless,
\begin{align}
\frac{1}{\Gamma} \frac{d\Gamma(W_L \to f_1 f_2)}{d\cos\theta^*} & \propto 1-\cos^2\theta^* \nonumber \\
\frac{1}{\Gamma} \frac{d\Gamma(W_T \to f_1 f_2)}{d\cos\theta^*} & \propto (1\pm\cos\theta^*)^2 \nonumber
\end{align}
where $\pm$ refers to the two possible transverse polarizations and the angle $\theta^*$ is defined with respect to the direction of the $W^\pm$'s motion in the lab frame. Higher order corrections will disrupt this pattern, but the effect has been shown to be small~\cite{Groote:2012xr}. The two distributions are shown below in Fig.~\ref{fig:partonlevel}.
\begin{figure}[!ht]
	\centering
	\includegraphics[width=0.7\textwidth]{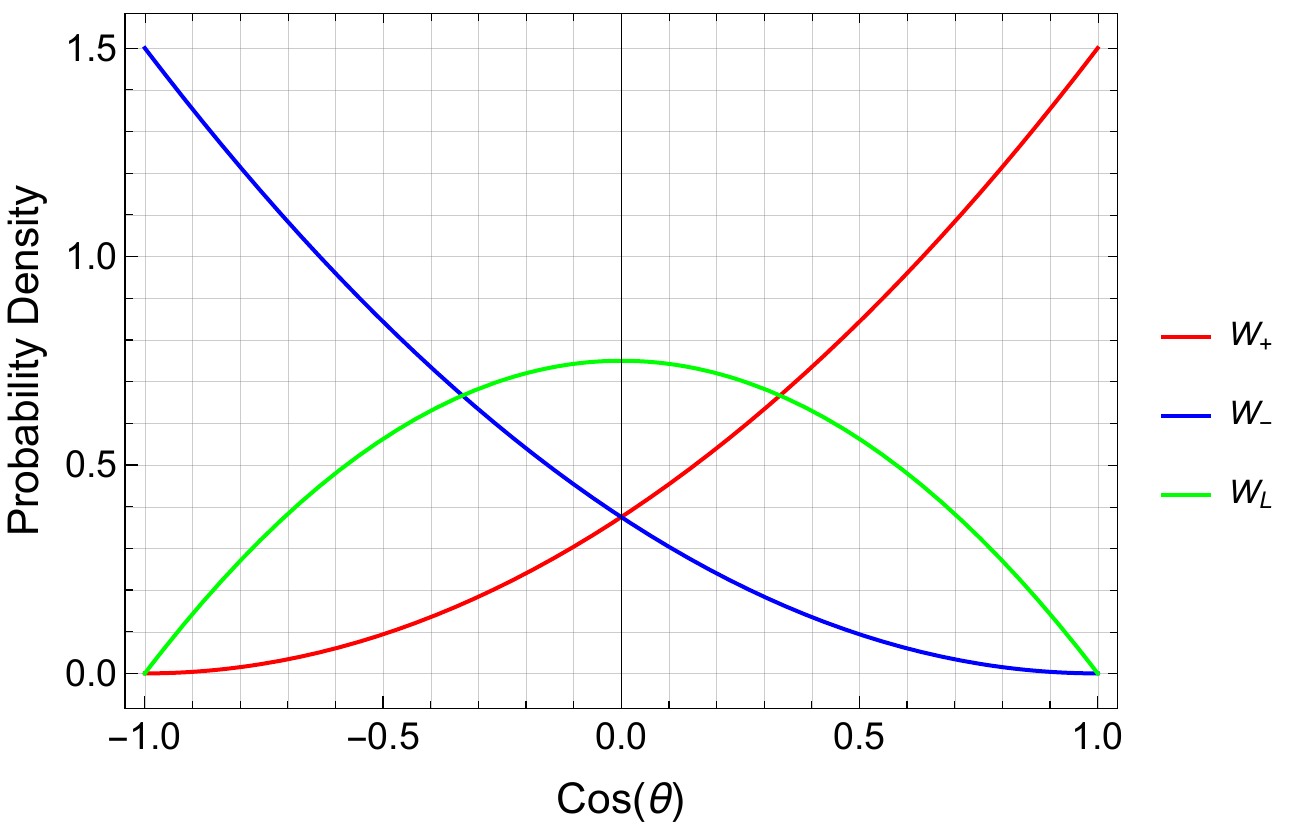}
	\caption{Parton level angular distribution. Analogous relations can be derived for $Z$ bosons. However, as the $Z$ couples to both left and right-handed fermions, the relations are not as simple and the distributions not as distinct as the $W^\pm$ case.}
	\label{fig:partonlevel}
\end{figure}
The $\cos{\theta}^*$ information can also be captured in the lab frame as $p_\theta \equiv \frac{\Delta E}{p_V}$, where $\Delta E$ is the energy difference between the decay products and $p_V$ is the momentum of the $W^{\pm}/Z$~\cite{De:2020iwq}.

For $Z$ bosons, the leptonic modes allow clear access to the polarization info, though at the price of a small branching fraction. For leptonic $W^\pm$ events where there is only one neutrino (i.e $W(\ell\nu) + \text{jets}$ or $W(\ell\nu)W(jj)$), one can attribute all missing energy in the event to the neutrino and solve for the longitudinal neutrino momentum by requiring the `neutrino' and the charged lepton reconstruct the $W^\pm$. This method yields the full lab-frame neutrino four-vector,  but it automatically introduces a two-fold ambiguity, as the $W^\pm$ mass constraint is quadratic, and is subject to uncertainties from mis-measured missing energy and $W^\pm$s that are slightly off-shell.

For hadronically decaying $W^{\pm}/Z$, there is no clean solution due to the usual difficulties of jet physics -- mis-measurement and the challenges of correctly filtering the $W^\pm/Z$-decay quark/anti-quark (and their associated radiations) from hadronic activity unrelated to the $W^\pm/Z$.  Any uncertainties in extracting the momenta of the $W^{\pm}/Z$ or their decay products results mixes the polarizations and makes them harder to separate. For example, boosts along the true $W^\pm/Z$ momentum don't mix polarizations, i.e. a transversely polarized $W^\pm$ are left invariant by longitudinal (along the direction of motion) boosts, and the longitudinal polarization remains longitudinal, but this is not the case for boosts along other directions, such as the direction of an incorrectly reconstructed $W^\pm/Z$. Due to these complexities, polarization studies have focused primarily on leptonic $W^\pm/Z$.

One recent exception is Ref.~\cite{De:2020iwq}, which studied the polarization of boosted hadronic $W^\pm$ by using jet substructure techniques to extract $p_\theta \equiv \frac{\Delta E}{p_V}$. Specifically, by using the variable $N$-subjettiness~\cite{Thaler:2010tr}, the fat $W$ jet gets factored into smaller pieces, and can be used to select out clusters of energy to serve as proxies for the underlying quark and anti-quark. Bolstered by techniques to clear away extraneous QCD radiation~\cite{Ellis:2009su}, the subjets faithfully represent the partonic physics, and the authors demonstrate polarization discrimination in vector boson fusion and in the presence of a hypothetical new resonance that decays to $W^+W^-$. The price for the substructure approach is additional cuts -- on the mass of the $W$ jet, the mass fraction of the subjets, and the `subjettiness' variable itself. These lead to a more accurate sample, but reduce the number of events and can potentially reintroduce interference among the different polarizations~\cite{Ballestrero:2017bxn, Mirkes:1994eb, Stirling:2012zt, Belyaev:2013nla}. We would like to study in the polarization differences using the same physics -- the trace of the angular/energy correlations left in the $W$-jet substructure -- but using a more inclusive, though arguably less transparent, method. 

Before describing our method, it is useful to quantify how well one can possibly differentiate polarizations, e.g. as if we were able to work at parton level. For hadronic $W^\pm$, we cannot separate $W^+$ from $W^-$, so we must combine them. Let us introduce $|\cos{\theta}_{cut}|$, and classify all events with $|\cos{\theta^*}| \le |\cos{\theta}_{cut}|$ as longitudinal. Varying $\cos{\theta}_{cut}$, we trace out a curve in efficiency vs. mistag rate. This curve is shown in Fig.~\ref{fig:roc}, with the efficiency axis  labeled "true positive rate" and the mistag rate as "false positive rate" to make the connection with our later network results easier. Tracing that curve, we see about $60\%$ longitudinal $W$ can be successfully identified with a fake rate of $30\%$, or $80\%$ success with a $50\%$ fake rate. These partonic, and therefore `best case', efficiency/fake rates are fairly poor, especially when compared to rates from top/Higgs/$V$ `taggers' that differentiate between massive objects and QCD~\cite{Butter_2018, Barnard_2017, Lim_2018}. This is not surprising, given that we are aiming to distinguish between $W^\pm$s that have the same gross kinematic features ($p_{T,W}, \eta_W$) yet differ in polarization, so the angle between decay products is our only handle and the populations have non-negligible overlap near $\cos\theta^* \sim 0.5$. 


\section{From parton level to particle level: network setup and training}
\label{sec:network}

Moving from parton level to more realistic, detector level signals, we will attack this problem using jet images and deep neural networks (DNN). Deep neural networks have been shown to be a powerful tool for discriminating among different particles, such as quark vs. gluon jets~\cite{gluejet1,PETERSON1994185}, $W^\pm$ vs. QCD \cite{Barnard_2017} or tops vs. QCD~\cite{Macaluso_2018, Butter_2018}, displaying superior performance over analyses using kinematic variables alone.  Among different networks, we focus on convolutional neural networks (CNNs) which take boosted jet images as input and allow on the network to pick up on minute angular and energetic correlations among jet constituents that are inherited from the initial partons. In the following sections, we describe the network construction, image preprocessing, and supervised training samples, then present visualizations of the trained network's performance and predictions. 

\subsection{Monte Carlo tools and jet preprocessing procedure}
\label{sec:preproccess}

To simulate boosted $W^{\pm}$ bosons, we use {\tt MadGraph5v2.6.5} and {\tt MadGraph5v2.7.0}~\cite{Alwall:2014hca}\footnote{The recent updates on \texttt{MadGraph} allows us to generate polarization enforced events} at leading order and a center of mass energy  of 13 TeV. The parton level events are fed through {\tt PYTHIA}~\cite{Sjostrand:2014zea,Sjostrand:2006za} to incorporate showering and hadronization, then through {\tt Delphes}~\cite{deFavereau:2013fsa} to add detector effects. From the {\tt Delphes} calorimeter output, we extract a list of all charged and neural particle four-vectors in the event.\footnote{Specifically, we use the {\tt EFlowTrack} {\tt Delphes} branch for charged particles, the {\tt EFlowNeutralHadron} branch for neutral hadrons, and the {\tt EFlowPhoton} branch for photons.} The list of four-vectors is clustered into `fat' jets via {\tt FastJet}\cite{Cacciari:2005hq,Cacciari:2011ma} using the anti-kT algorithm with $R = 1.0$, minimum $p_T = 100\, \gev$ and max $|\eta| = 2.5$\footnote{We use the package {\tt Pyjet}\cite{pyjet} as a wrapper for {\tt FastJet}}. These jets are pre-processed and pixelized, and the pixels used as the inputs to our neural network. Preprocessing formats the jets, centering them and minimizing any angular anisotropy, so extraneous features are not picked up by the network to distinguish between samples. We follow the preprocessing steps from Ref.~\cite{Barnard_2017}~\footnote{In addition to centering and rotating, Ref.~\cite{Barnard_2017} also zooms, or rescales the $p_T$ of the image constituents so they can view jets across a wide range of $p_T$. As we focus on boosted $W$ jets in a few, relatively small $p_T$ windows, we do not perform this step.}:
\begin{enumerate}
\item[1.)] Re-cluster the fat jet into subjets using the Cambridge/Aachen algorithm~\cite{Dokshitzer:1997in,Wobisch:1998wt} with $\Delta R = 0.3$ and minimum $p_T = 1\,\gev$.
\item[2.)] Translate jet constituents' $(\eta,\phi)$ position to put the highest $p_T$ of leading subjet at the origin. 
\item[3.)] Rotate all jet constituents so that the highest $p_T$ of sub-leading subjet is located below the origin. 
\item[4.)] Reflect based on the number of subjets. For 2 subjets in a clustered jet, sum over $p_T$ of left and right side of the image to place higher $p_T$ sum on right hand side. For 3 or more subjets, reflect the jet image so that the third leading subjet is located on the right hand side of the image. Events with only one subject are rejected.
\end{enumerate} 
Next, the formatted jets are pixelized in $\eta,\phi$ in a $20 \times 20$ grid, with each direction spanning $-1$ and $1$ around the center of the fat jet, keeping with the pixel size used in Ref.~\cite{Barnard_2017}. The value of each grid point in the accumulated  $p_T$ value of the particles in that square.

\begin{figure}[!ht]
	\centering
	\includegraphics[width=0.45\textwidth]{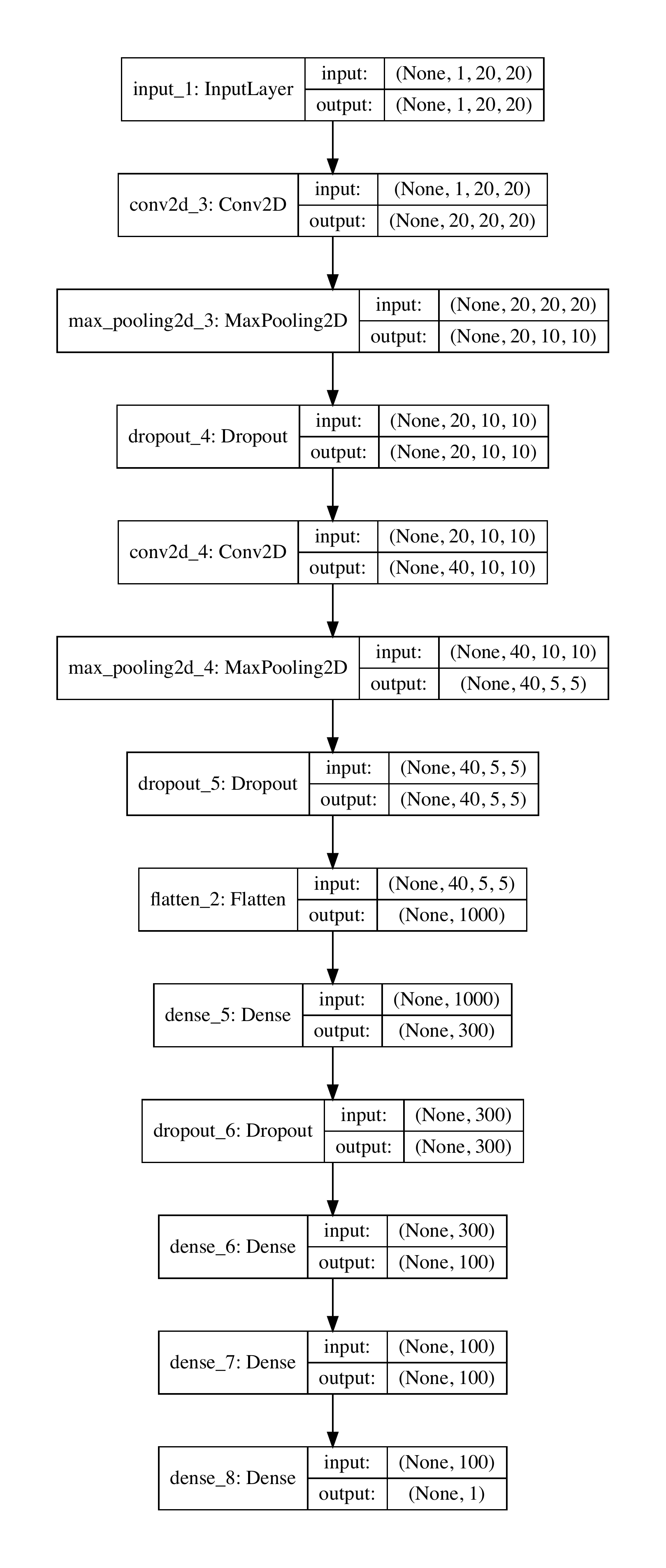}
	\caption{Visualization of CNN structure}
	\label{fig:network}
\end{figure}

\subsection{Neural Network structure and event information}
\label{sec:nnstruct}

The pixelized $20\times 20$ jet images form the first layer of our convolutional neural network (CNN). After the input layer, we follow the typical CNN structure example\footnote{Detailed Keras sample code can be found at https://www.tensorflow.org/tutorials/images/cnn} provided by Keras \cite{chollet2015keras} of a combination of convolutional and fully connected dense layers. Specifically, the input layer is followed by 2 dimensional convolutional layer with 20 kernels of size 4. This layer is subject to Max pooling with Dropout, then fed into a second convolutional layer with 40 kernels of size 4. Max pooling with Dropout is followed by the second convolutional layer output. After again pooling with Dropout, the result is then flattened and fed into 3 dense layers with 100 units each. Finally, the last dense layer it is connected to the output layer of 1 unit. Throughout the network, we use the rectified linear unit (ReLU) function to introduce non-linearity, except for the output layer which has a sigmoid activation function. With this architecture, the output sits in the range $[0,1]$ and can be interpreted as the probability that a given even comes from a longitudinal $W$. The structure of the our network is illustrated in Fig.~\ref{fig:network}.

We arrived at this network architecture and set of hyperparameter parameters by optimizing run time and performance on training samples (to be discussed shortly). In addition to the CNN, we explored how two networks from the literature performed. The networks we tested are MaxOut~\cite{Barnard_2017} and ResNet~\cite{he2015deep}, with structure displayed in Fig.~\ref{fig:maxout}, \ref{fig:resnet} respectively. MaxOut is a fully connected dense network with dedicated layers designed to mimic the filter features of a CNN and was built with the goal of differentiating $W$-jets from QCD, while ResNet (short for Residual Network) is an image based network that contains skipped connections in an effort to avoid vanishing gradient issues; it is significantly more complicated than our CNN. Comparing with MaxOut is a useful cross check with previous literature~\cite{Oliveira_2016}, while comparing with ResNet illustrates whether a more advanced network architecture is worth the added number of parameters. The results from these networks along with more details of their architecture and how it differs from the CNN we use are presented in Appendix \ref{appendix}.

\subsection{Network Training}
\label{sec:nntrain}

As our training (MC) samples, we want processes that have pure $W^\pm$ polarization. For transverse $W^\pm$ bosons, 
$pp \to W^\pm + \text{jets}$ is an easy choice, while for longitudinal $W^\pm$ we use $pp \to H \to W^+W^-$, where $H$ is a fictitious heavy Higgs boson with mass $500\, \gev$~\footnote{To generate this signal, we use the HEFT model included within MadGraph.}. An alternative sample of longitudinal $W^\pm$ that could be used more readily in a data-driven approach is associated production $pp \to hW$.  For all training samples we lump hadronic $W^+$ and hadronic $W^-$ events together, as they are experimentally indistinguishable. 

We also break up the training into two $p_T$ bins: $p_T \in [200\,\gev, 300\,\gev]$, which we will refer to as the `low-$p_T$' sample, and  $p_T \in [400\,\gev,500\,\gev]$, the `high-$p_T$' sample. These choices are motivated by the fact that,  $W^{\pm}$ with $p_T \le 200\, \gev$ are not boosted enough for their decay products to fall within $\Delta R \le 1$  (our fat jet definition), while $W^\pm$s with $p_T > 500\, \gev$ suffer from a low rate and tend to have such collimated decay products that both end up with the same subjet. Importantly, we do impose any cuts other than $p_T$. This can be contrasted with Ref.~\cite{De:2020iwq}, where additional substructure cuts, such as mass drop and N-subjettiness must be applied to `locate' the primary $W$ decay products needed in $p_\theta$. These additional cuts have a signal efficiency of $\mathcal O(50\%)$\footnote{Ref.~\cite{De:2020iwq} considered slightly different $p_{T,W}$ regions for their analysis, $p_{T,W} \sim 800-1000$ GeV, and it is possible the efficiencies for the additional substructure cuts carry some $p_T$-dependence.}, thereby reducing the event sample size and ultimately feeding into the uncertainty.\footnote{At this point we are assuming the same starting point as Ref~\cite{{De:2020iwq}} -- a sample of pure $W$s of unknown polarization. A more accurate comparison requires including non-$W$ backgrounds. We will discuss the role of other backgrounds a little in Sect.~\ref{sec:results}, deferring a more complete study to later work.}

As we create the training and validation samples, it is crucial to keep the number of events for each polarization the same to avoid unequal trainings.  As a result, we use 340k for training and 85k for validating at lower $p_T$. At higher $p_T$, we use 236k for training and 59k for the validation. When training the network, we intervene and stop if there is no significant enhancements for 10 iterations within maximum 200 epochs\footnote{ For other hyperparameter settings: we use Keras callback EarlyStopping with patience = 15 and ReduceLROnPlateau with patience = 5} on the validation set.  The training/validation samples and their network output are summarized below in Table~\ref{tab:trainingfo}.
\begin{table}[h!]
	\centering
	\begin{tabular}{|c|c|c|}\hline
	$p_T$ range & Number of training/validation events & validation accuracy  \\ \hline
	$200\, \gev \le p_T \le 300\, \gev$ & 340k & 63\% \\
	$400\, \gev \le p_T \le 400\, \gev$ & 236k & 64\% \\ \hline
	\end{tabular}
	\caption{Summary of training samples and network validation accuracy. For both $p_T$ bins, we use 20\% of the sample for validation and 80\% for training.}
\label{tab:trainingfo}
\end{table}

\begin{figure}[!h]
	\centering
	\includegraphics[width=0.55\textwidth]{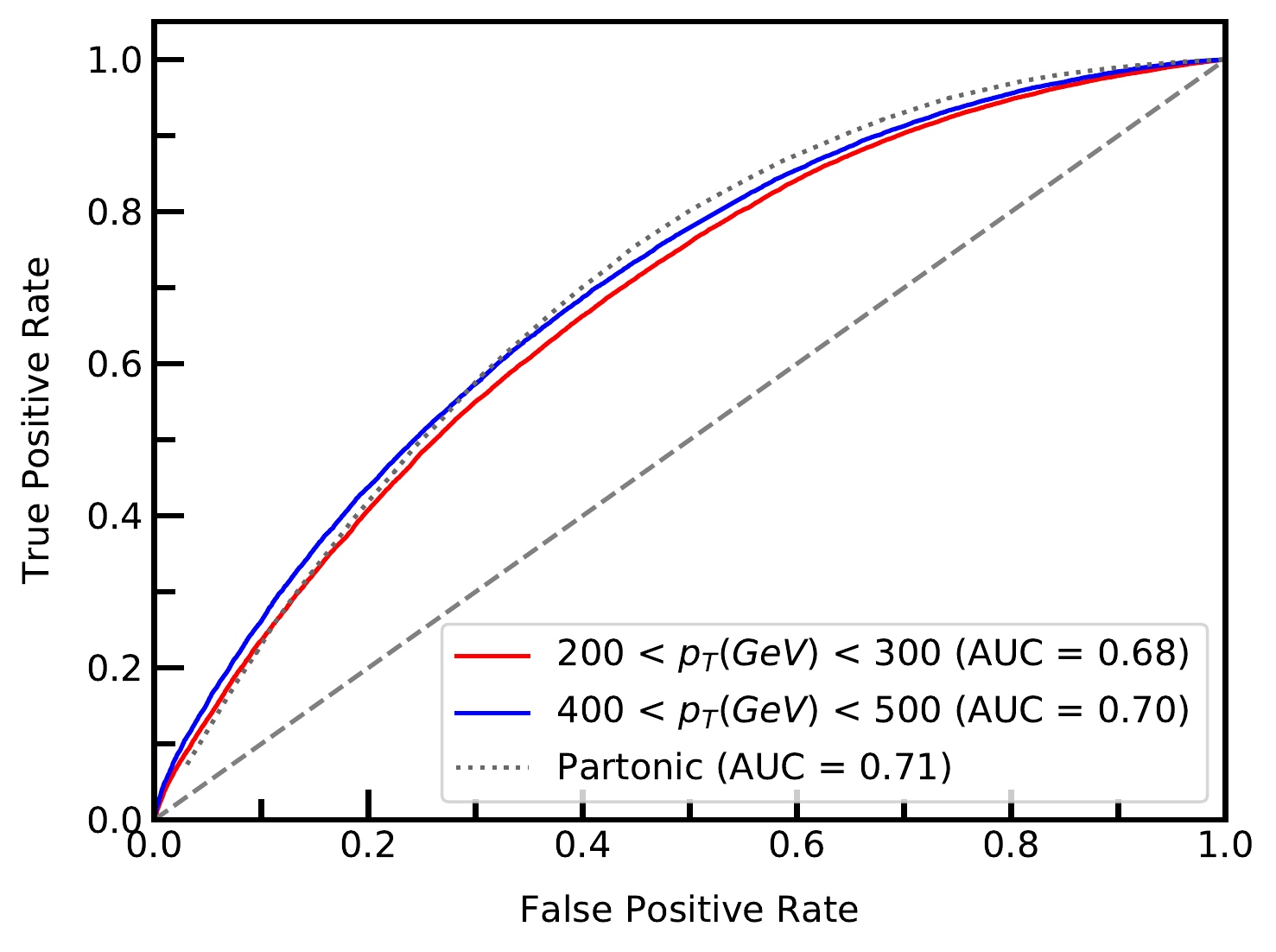}
	\caption{Receiver Operating Characteristic Comparison:
	Partonic ROC curve is shown as a reference line to compare with each trained network's performance comparison. As the curve bends more toward the upper left corner, the performance of the network increases. Considering the angular distribution as the theoretical limit, our networks for both $p_T$ bins shows the clue of reaching the limit.}
	\label{fig:roc}
\end{figure}

We plot the Receiver Operating Characteristic (ROC) curve based on the validation samples in order to visualize the network's performance. In Fig.~\ref{fig:roc}, the partonic curve indicates the theoretical maximum of the training calculated in the previous section, and we observe that our trained networks for both $p_T$ samples nearly matches to the partonic version. While it is good to see that the network approaches the ideal/partonic curve, the true positive rates are not significantly larger than the corresponding false positive rates. As such, event-by-event tagging using our network is not particularly powerful. This result is seconded by the  network's accuracy, $\sim 60\%$, defined as the correct classification probability when the threshold (value between 0 and 1 where we classify the event as longitudinal or transverse) is set to 0.5.  Therefore, instead of treating the network as a variable to cut on, event by event, to select a certain polarization population, we will keep all events and use the network output of the entire ensemble to extract the polarization fraction.\footnote{For another example using ML event ensembles to extract information about model parameters (though with a DNN and engineered variables rather than a CNN and images), see Ref. \cite{flesher2020parameter}.}

\begin{figure}[!h]
	\centering
	\begin{subfigure}[b]{0.43\textwidth}
		\centering
		\includegraphics[width=\textwidth]{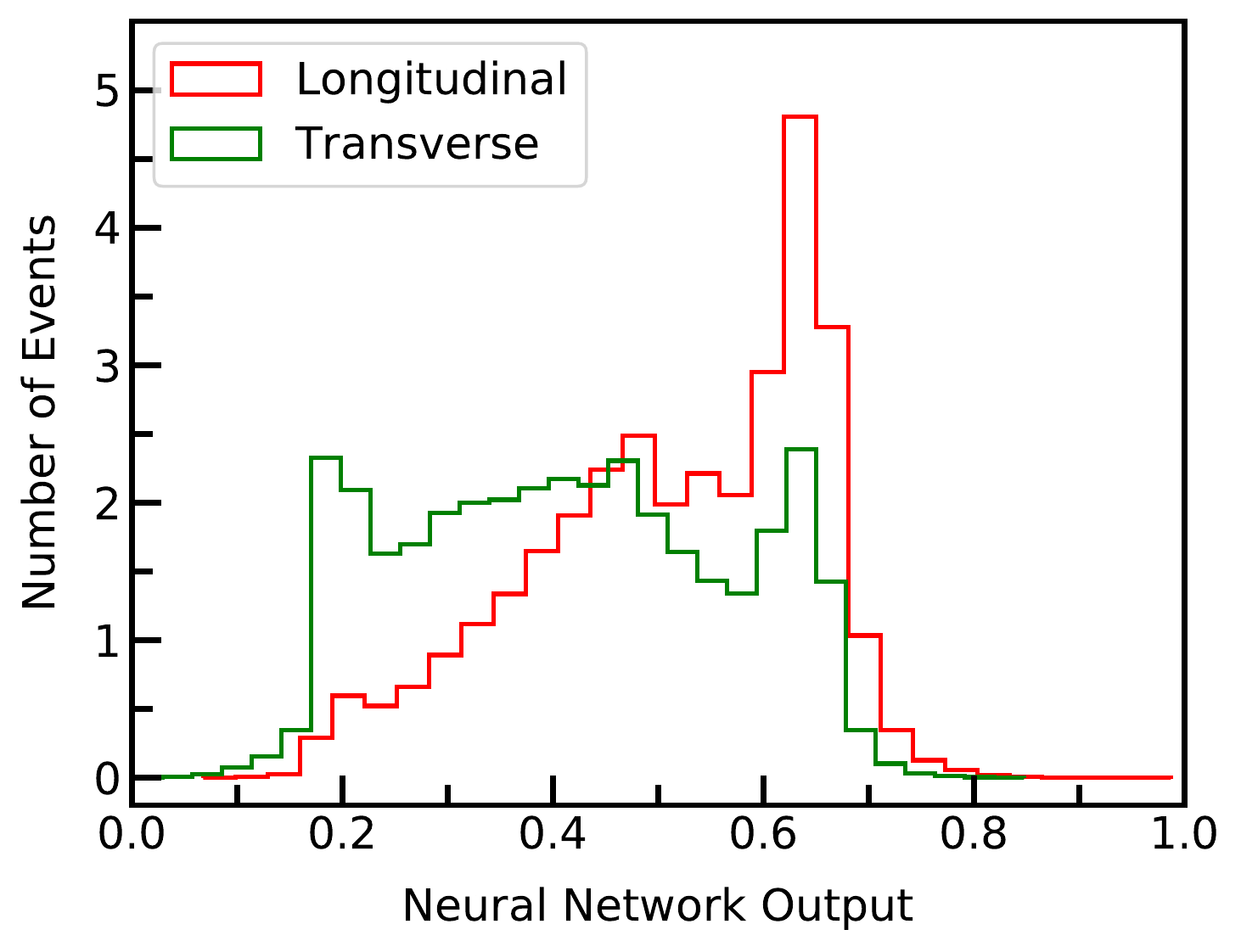}
		\caption{$200 \le p_T \le 300$}
	\end{subfigure}
	\hfill
	\begin{subfigure}[b]{0.43\textwidth}
		\centering
		\includegraphics[width=\textwidth]{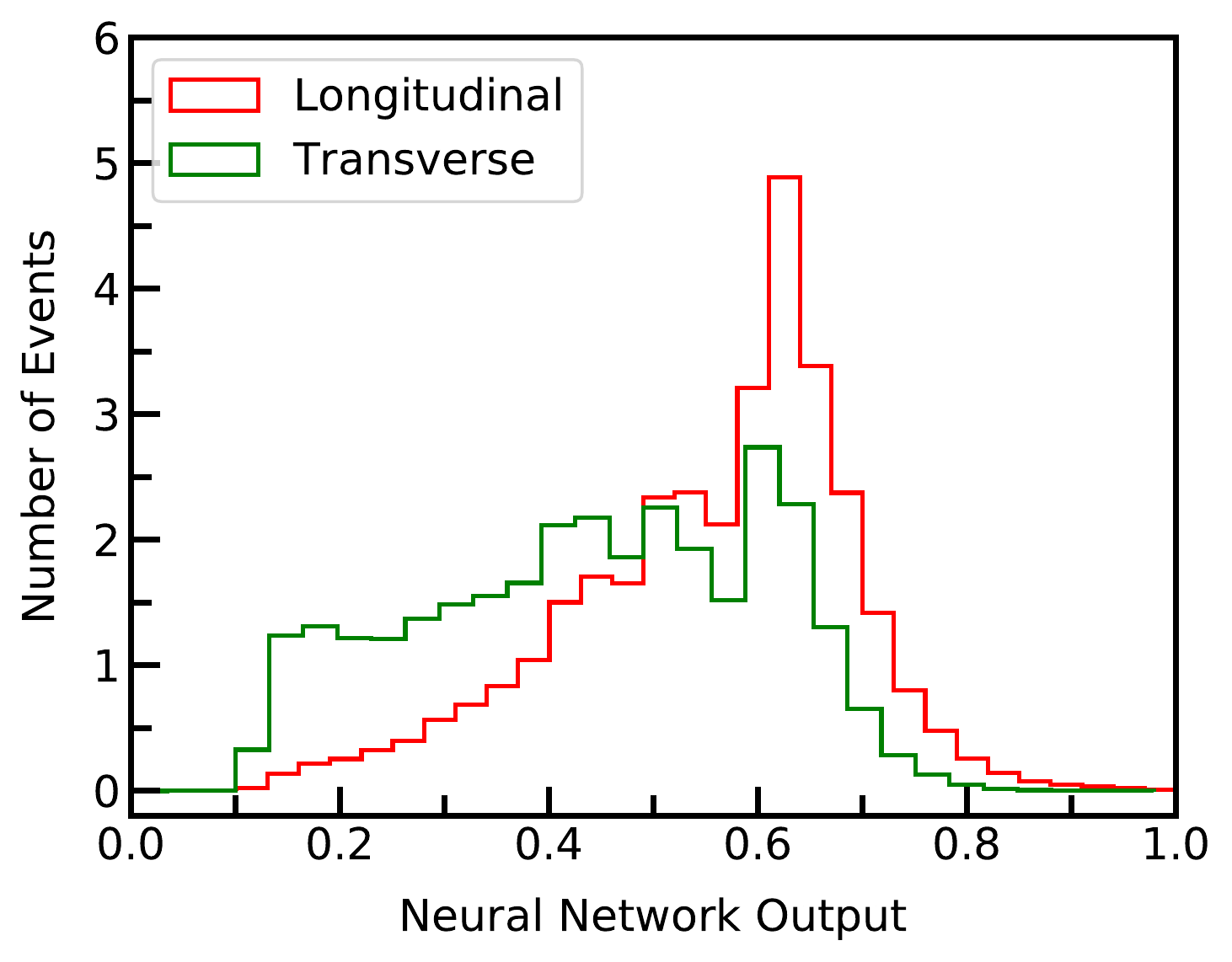}
		\caption{$400 \le p_T \le 500$}
	\end{subfigure}
	
	\caption{The distribution of network outputs for two different $p_T$ bins, determined using our validation dataset. The distribution for the longitudinal sample ($H \to W^+W^-)$ is shown in red and peaks near 1, while the green line shows the transverse $W$ sample distribution (from $W + \text{jet}$) and has more support towards 0.  The distributions are unit normalized.}
	\label{fig:outputdistrb}
\end{figure}

The (area normalized) network output for the transverse ($pp \to W^{\pm}+\text{jets}$) and longitudinal ($pp \to H \to W^+W^-$) validation samples are shown below in Fig. \ref{fig:outputdistrb} for the two $p_T$ regions.  We can identify several features in the distributions: a true peak, a false peak and a central region. The true peak corresponds to when the network properly classifies a validation event, the false peak represents the mistagging of the network and tends to coincides with the true peak of true positive events, and the central region is populated by obscure outputs. Obviously, both the false peak and central regions contribute to diluting the performance.  Comparing the two $p_T$ regions, the network output in the higher $p_T$ sample has a larger fluctuation in the central region.  

Knowing the network templates for the purely longitudinal and transverse samples, we interpolate between them to fit the network output from a signal whose polarization composition we'd like to find. Specifically, we interpret the network output as a probability distribution ($D_i(x)$) and set.
\begin{equation}
	f_L\times D_L (x) + f_T \times D_T (x) = D_\text{unknown} (x)
\end{equation}
Here $f_L, f_T$ are the longitudinal and transverse fractions and $D_L(x), D_T(x)$ are network distributions determined from the validation sets. Setting $f_T = 1 - f_L$, multiplying by $x$ and integrating, we find a relation between the expectation values of the validation distributions and the distribution with unknown polarization composition.
\begin{equation}
f_L \left<x_L\right> + (1- f_L) \left<x_T\right> = \left<x_\text{unknown}\right>
\end{equation}
Solving for $f_L$, we find:
\begin{equation}
f_L = \frac{\left<x_\text{unknown}\right>- \left<x_T\right>}{\left<x_L\right>- \left<x_T\right>}
\label{eq:xval}
\end{equation}
 

\section{Results}
\label{sec:results}

\subsection{Polarization analysis on SM $pp \to W^\pm Z$}
\label{sec:SMWZ}

Our first test case is SM $pp \to W^\pm(jj) Z(\ell\ell)$. This SM process is a good test candidate since it has a relatively high cross section and is not dominated by a single polarization (unlike, for example, $pp \to W^{\pm}H$ which is completely dominated by longitudinal $W$s); $pp \to W^\pm(jj) Z(\ell\ell)$ also experimentally clean, as the presence of the leptonic $Z$ will mitigate backgrounds from top quark production -- a handle we don't have if looking at $pp \to W^{\pm}(jj)W^{\mp}(\ell \nu)$. Using a newly introduced feature of {\tt MadGraph5v2.7.0}, we can specify the $W/Z$ polarization when generating events. This lets us quickly check the truth-level polarization fraction for each set of cuts. 

We generate 1M testing events $pp \to W^\pm Z$ samples for lower $p_T$ bin (20k for higher $p_T$ bin), following the same preprocessing as training/validation samples.  In situations where there are multiple jets passing the kinematic criteria, we select the jet whose $\phi$ coordinate is closest to $-\phi$ of the reconstructed $Z$. The testing events play the role of the sample with unknown polarization composition in the discussion above, and size of the samples we generated is related to the number of expected events at the end of the HL-LHC era, as we will explain. Running these events through our network, then fitting the network output to a sum of the longitudinal and transverse templates, we find $f_L$.  The results are quantified in Table \ref{tab:SMWZresult}, with the output average method showing good agreement with the ideal values.
\begin{table}[!h]
	\centering
	\begin{tabular}{|c|c|c|c|}
		\hline
		$p_T$ range  & $\sigma(p p \to W^{\pm}(jj)Z(\ell \ell))\, $ (fb) &  truth $\sigma_{L}/\sigma_{\text{tot}} $  & predicted $f_L$\\ \hline
		$200\,\gev\, \le p_T\, \le 300\, \gev$ &6.67& 0.265     & 0.259  $\pm $ 0.013       \\ \hline
		$400\,\gev\, \le p_T\, \le 500\, \gev$ &0.35& 0.304    &  0.300  $\pm$  0.033    \\ \hline
	\end{tabular}
	\caption{Longitudinal polarization fraction comparison between truth and using the network output average. The truth value of $f_L$ is calculated using cross section provided by \texttt{MadGraph}.  The cross section shown in the second column includes branching fractions ($\ell = e,\mu$) and the acceptance for the $p_T$ cuts for each row. For the parton level cuts and jet requirements we have assumed, the acceptance cut efficiency is 59\% for the lower-$p_T$ sample and 65\% for the high-$p_T$ sample. The uncertainty on the extracted $f_L$ is determined using the bootstrap method explained in the text. }
	\label{tab:SMWZresult}
\end{table}

The network output values in Table~\ref{tab:SMWZresult} include uncertainty bands, which were estimated using the following approach:
\begin{itemize}
\item We assume that the statistical uncertainty on $\langle x_T \rangle$ and $\langle x_L \rangle$ are small as they can be determined by large simulated datasets. Had we carried out a binned analysis rather than working with the network average, this assumption would have been hard to justify given our total training sample size of $\sim$ few hundred thousand events.
\item Assuming that the uncertainties are uncorrelated, propagation of uncertainty leads to
\begin{equation}
\label{fig:fLerror}
\sigma_{f_L}^2 = \left( \frac{\partial f_L}{\partial \left<x_\text{unknown}\right>}\right)^2\sigma_{\left<x_\text{unknown}\right>}^2 = \left( \frac{1}{\left<x_L\right>-\left<x_T\right>}\right)^2\sigma_{\left<x_\text{unknown}\right>}^2
\end{equation}

\item For a single set of testing data, we only get one number -- the network average. To determine the uncertainty on the network average we can run pseudo-experiments (`bootstrapping' technique, in network terminology ). Specifically, we randomly select subsets of the testing data that correspond to the number of signal events expected for a given luminosity, and calculate the network output for this subset. Iterating this procedure, we can use the distribution of results to define the uncertainty.

\end{itemize}

For this particular example, we select the size of the $pp \to W^{\pm}Z$ dataset to correspond to the number of $pp \to W^{\pm}Z$ events at the end of the HL-LHC run. Using the (LO) cross section from Table~\ref{tab:SMWZresult} and assuming $\mathcal L = 3\,\text{ab}^{-1}$, this corresponds to 20k events for $p_T \in [200\,\gev, 300\,\gev]$ and 1k for $p_T \in [400\,\gev, 500\,\gev]$. Iterating 20 times, and plugging the extracted $\sigma^2_{\langle x_{unknown} \rangle}$ into Eq.~\eqref{fig:fLerror}, we find the uncertainty on $f_L$ quoted in the last column of Table~\ref{tab:SMWZresult}.  We find that $\sigma^2_{\langle x_{unknown} \rangle}$ does not depend strongly on the number of iterations, provided the number is $\gtrsim$ few.  If we instead use batches corresponding to event sizes for $300\, \text{fb}^{-1}$ ($2$k events for  $p_T \in [200\,\gev, 300\,\gev]$, 100 events for $p_T \in [400\,\gev, 500\,\gev]$), the uncertainty on $f_L$ increases to $0.033$ ( $p_T \in [200\,\gev, 300\,\gev]$) or  0.132 (for $p_T \in [400\,\gev, 500\,\gev]$); for event sizes corresponding to $150\, \text{fb}^{-1}$, the uncertainty becomes $0.044$ (0.190) for the low (high) $p_T$ bins respectively. 

Looking at Table~\ref{tab:SMWZresult}, we see that the network prediction reproduces the truth value. Based off of our pseudo-experiment test, the uncertainty on $f_L$ for $p_T \in [200\,\gev, 300\,\gev]$ is $\sim 1.2/\sqrt{N_{events}}$ for the number of events available with the  (roughly) the current LHC luminosity ($150\, \text{fb}^{-1})$, rising to $1.8/\sqrt{N_{events}}$ for $3\, \text{ab}^{-1}$.
 
Of course, the numbers quoted above assume we have been handed a sample of pure $pp \to W(jj)Z(\ell \ell)$ events and therefore ignores the presence of other SM backgrounds. As our study here is simply a first step in hadronic $W$ polarization analysis, we will stick with idealized `$WZ$-only' events for the remaining examples. However, it is worthwhile to consider  how other backgrounds will impact our story. For a $pp \to W^{\pm}(jj)Z(\ell\ell)$ signal, the main worry is $pp \to Z(\ell\ell) + \text{jets}$\footnote{Other backgrounds are present, such as fully leptonic $\bar t t$ production and $ZZ/Z\gamma$, however they are smaller; leptonic $\bar tt$ can be suppressed by the requirement of an on-shell leptonic $Z$, while $ZZ/Z\gamma$ have small production rates.}. There has  been lots of recent progress distinguishing massive vector bosons from QCD, both with substructure analysis and jet images~\cite{Barnard_2017}. The degree to which that background impacts our quantitative results depends on the W-tagging algorithm. As a back-of-the envelope calculation, an additional cut to filter out QCD with efficiency epsilon will inflate the uncertainty on our polarization fraction extraction by $\sqrt{ 1/\epsilon}$. This estimate ignores any biases the QCD-vs.-$W$ cuts introduce, or pollution from mistags. As an example, the $W^\pm$ tagger in Ref.~\cite{Barnard_2017} quotes a tagging efficiency of $\epsilon \sim 50\%$ for a fake rate of $7\%$, resulting in a $\sim 50\%$ inflation in the uncertainties from decreased signal statistics alone. Further work combining polarization analysis into existing $W^\pm$ vs. QCD algorithms and including all backgrounds would be interesting to pursue.
 
\subsection{Polarization analysis for dimension-6 operators}
\label{sec:SMEFTresults}

Having tested our method, we now explore how our well our polarization analyzer performs at detecting the presence of higher dimensional operators. Different operators contribute to different $W^{\pm}/Z$ gauge boson polarizations, therefore including them in processes involving electroweak gauge boson production can potentially change the ratio of transverse to longitudinal bosons.

There are several reasons to study this example. First, it is insensitive to the UV setup, as it can be applied to any scenarios one can map into the SMEFT framework. This can be contrasted with a test that assumes a particular UV content, i.e. a resonance. Second, while measuring the cross section is an obvious way to look for the presence of higher dimension operators, it's possible for new physics to have negligible impact on the cross section, either because coefficients are small or because different effects conspire and cancel. In these cases, analyzing the polarization provides another handle and can potentially spot new physics or disentangle effects that the cross section is blind to.

We will focus on two particular higher dimensional operators that can impact the process $pp \to W^{\pm}Z$: 
\begin{align}
\mathcal L_{NP} = c_{W}\mathcal O_W + c_{3W}\mathcal O_{3W}
\end{align}
where $c_{W, 3W}$ are dimensionless Wilson coefficients and 
\begin{align}
\mathcal O_W &= \frac{i\,g}{m^2_W}(H^{\dagger}\sigma^a \overleftrightarrow{D}^\mu H) D^\nu W^a_{\mu\nu} \\
\mathcal O_{3W} &= \frac{i\,g^3}{m^2_W}\, \epsilon_{abc} W^{a\nu}_\mu W^b_{\nu\rho} W^{c\rho\mu},
\label{eq:NPops}
\end{align}
following the convention of Ref.~\cite{Alloul:2013naa}.  The factor of $m_W$ in the denominator is a bit unconventional, however since any measurement will only reveal information on the ratio of the Wilson coefficient $c_i$ to the scale suppressing the operator, we can always translate this normalization to any other suppression scale $\Lambda$. In our simulations, the vertices contained in $\mathcal O_W, \mathcal O_{3W}$ are allowed to enter a given amplitude/diagram once. As such, the cross section is a quadratic function of the Wilson coefficients $c_W, c_{3W}$. The linear term represents the interference between the SM and the higher dimensional operators, while the quadratic term contains the square of the new physics amplitudes.  Finally, as we are picking a subset of dimension-6 operators, this study should be viewed as a straw man to illustrate a technique rather than a genuine SMEFT analysis, as the latter requires working with a complete basis and a more consistent treatment of quadratic EFT effects.

From the field content of $O_W$ and $O_{3W}$, we suspect that $O_W$ will affect the production of longitudinal $W$ while $O_{3W}$ only includes field strengths and can therefore only participate in transverse production. This thinking is backed up by Ref~\cite{Liu_2019:dim6}, which analyzed diboson production in the presence of certain dimension-6 and -8 operators.\footnote{The impact of higher dimensional operators on the polarization breakdown can be found by studying how various $W^\pm_\lambda Z_{\lambda'}, \lambda, \lambda' = T, L$ subprocesses depend on the scales in the problem and identifying contributions that grow with the energy of the process. Following Ref.~\cite{Liu_2019:dim6},  the $W_L Z_L$ cross section contributions involving $c_W$ (both linear and quadratic) grow with energy, while $c_{3W}$ does not contribute, while for $W_T\, Z_T$ all effects involving $c_W$ are suppressed, the linear $c_{3W}$ term is constant, and the $c^2_{3W}$ contribution grows with energy.}

As a first test we turn on one operator at a time using Wilson coefficient value $10^{-3}$ for $c_W$ and $3 \times 10^{-3}$ for $c_{3W}$. Rescaling to operators suppressed by $\Lambda^2 = (1\, \tev)^2$ and with no explicit factors of $g$, this choice corresponds to an overall coefficient of $0.1$ for $(H^{\dagger}\sigma^a \overleftrightarrow{D}^\mu H) D^\nu W^a_{\mu\nu}$ and $0.13$ for $\epsilon_{abc} W^{a\nu}_\mu W^b_{\nu\rho} W^{c\rho\mu}$. For Monte Carlo purposes, we use the UFO implementation of $\mathcal O_{W}, \mathcal O_{3W}$ from Ref.~\cite{Alloul:2013naa}.  The network $f_L$ output using the method of Sect.~\ref{sec:network} for each operator choice and $p_T$ bin is shown below in Table ~\ref{tab:dim6WZresult}, along with the cross sections.

\begin{table}[!ht]
	\centering
	\begin{tabular}{|c||c|c|c|c|}
		\hline
		&                               $p_T$ range & $\sigma(pp \to W^{\pm}Z)$ (fb) & truth $\sigma_{L}/\sigma_{tot}$  &predicted $f_L$ \\ \hline
		\multirow{2}{*}{$O_W$} &  $200\, \gev \le p_T \le 300\, \gev$ & 6.93 & 0.311    & 0.297 $\pm$  0.010              \\ \cline{2-4} 
		& $400\, \gev \le p_T \le 500\, \gev$ &0.42 & 0.439    & 0.391  $\pm$  0.033             \\ \hline
		\multirow{2}{*}{$O_{3W}$} & $200\, \gev \le p_T \le 300\, \gev$ & 6.58 & 0.258    & 0.254   $\pm$ 0.011               \\ \cline{2-4} 
		& $400\, \gev \le p_T \le 500\, \gev$ & 0.50 & 0.198    & 0.181  $\pm$ 0.043               \\ \hline
	\end{tabular}
	\caption{Truth level and network average longitudinal fraction results when one dimension-6 operator at a time is included. As in Table~\ref{tab:SMWZresult}, the truth values were determined by restricting the $W$ polarization at generator level in \texttt{MadGraph}, and the quoted cross sections include branching ratios and acceptance for kinematic cuts. Uncertainties in the predicted $f_L$ are calculated using the method of Sec.~\ref{sec:SMWZ} and are based on pseudo-experiments with sample sizes matching the expected number of events for $3\,\text{ab}^{-1}$ of luminosity.}
	\label{tab:dim6WZresult}
\end{table}

Comparing our value of $f_L$ with the truth, we see that the network average performs well.  As expected, $\mathcal O_W$ impacts the longitudinal fraction, while $\mathcal O_{3W}$ impacts the transverse fraction. The sign of the impact depends on the sign of the Wilson coefficient and the relative size of the linear (in $c_W, c_{3W}$) and quadratic contributions to the cross section.

Using the uncertainties derived from pseudo-experiments with sample sizes corresponding to $3\, \text{ab}^{-1}$ of luminosity, we can take the ratio of the deviation in the polarization fraction (network $f_L$ value in the presence of the higher dimensional operator minus the SM value) to $\delta f_L$ as a rough measure of the discriminating power. We fine this ratio is: 3.2 for $O_W$, low-$p_T$, 4.1 for $O_W$, high-$p_T,$ 0.6 for $O_{3W}$, low-$p_T$ and $2.8$ for $O_{3W}$, high-$p_T$. The ratio is higher for $O_W$, despite the fact that $O_{3W}$ has a larger effect on the total cross section (for this benchmark point)\footnote{This is somewhat counterintuitive, given that $O_W$ impacts the longitudinal fraction and amplitudes with longitudinal vector bosons tend to grow with energy. However, the pieces in the amplitude that are quadratic in $c_W$, $c_{3W}$ grow with energy -- for all $W$ polarizations. These quadratic pieces, and the fact that $c_{3W} > c_W$ for this benchmark point, lead to a greater cross section changes from $O_{3W}$.}. If we divide the difference in total cross section (with operators versus SM) by $1/\sqrt{N_{events}}$ --  a proxy for the uncertainty on the cross section  -- we find much larger numbers, $\mathcal O(10)$. Therefore, at least for the benchmark values in Table~\ref{tab:dim6WZresult}, the total cross section is a more powerful measurement for detecting the presence of these operators. This is not surprising, as the polarization fraction is a more refined quantity.  However, the polarization fraction can provide insight into what type of operator is responsible for any observed change in cross section. For example, the difference in $f_L$ values in the presence of $O_W$ versus $O_{3W}$ for $200\, \gev \le p_T \le 300\, \gev$ in Table~\ref{tab:dim6WZresult} -- which have similar impact on the cross section -- is $O(4)$ times the full luminosity HL-LHC uncertainty on $f_L$. As with the SM study in Sec.~\ref{sec:SMWZ}, these numbers neglect the impact from processes other than $pp \to W(jj)Z(\ell\ell)$.

If we use smaller event samples to determine the uncertainty, corresponding to pseudo-experiments using smaller luminosity datasets, $\sigma_{f_L}$ increases. As an example, we find $\sigma_{f_L} = 0.035$ for $300\, \text{fb}^{-1}$ of luminosity ($\sigma_{f_L} = 0.047$ for $150\, \text{fb}^{-1}$ of luminosity) in the low-$p_T$ scenario for both $O_{W}$ and $O_{3W}$. Propagating these larger uncertainties through, we find the difference in polarization fraction between samples with $O_W$ and samples with $O_{3W}$ (for the values in Table~\ref{tab:dim6WZresult} ) is roughly $2$ ($1.2$) times $\sigma_{f_L}$.

As a second test, we explore a scenario where both $c_W$ and $c_{3W}$ are non-zero, but they have been tuned so that their net effect on the cross-section is negligible. This test examines how well the polarization breakdown works as a way to detect the presence of new physics, given no hints of anything BSM from the cross section alone. We adjust the size of both coefficients for each $p_T$ bin respectively: $c_W = -1.0\times 10^{-3}, c_{3W} = 5.0\times 10^{-3}$ for $p_T \in [200\,\gev, 300\,\gev]$ and $c_W = -1\times 10^{-4}, c_{3W} = 5\,\times 10^{-4}$ for $p_T \in [400\,\gev, 500\,\gev]$
\begin{table}[!ht]
	\centering
	\begin{tabular}{|c|c|c|c|}
		\hline
		$p_T$ range    & $\sigma(pp \to W^{\pm}Z)$ (fb) & truth $\sigma_{L}/\sigma_{tot}$ & predicted $f_L$ \\ \hline	
		200 GeV $\le p_T \le$ 300 GeV & 6.68                         & 0.202                           & $0.207 \pm 0.011$     \\ \hline
		400 GeV $\le p_T \le$ 400 GeV & 0.34                          & 0.285                           & $0.282 \pm 0.044$     \\ \hline
	\end{tabular}
\caption{Longitudinal fraction results, truth versus network average, in a scenario where both $c_W$ and $c_{3W}$ are nonzero. The coefficients have been chosen so there is essentially no impact on the $pp \to WZ$ cross section, which can be verified by comparing the second column here to the second column of Table~\ref{tab:SMWZresult}. Uncertainties in the predicted $f_L$ are calculated using the method of Sec.~\ref{sec:SMWZ} and are based on pseudo-experiments with event sizes matching the expected number of events for $3\,\text{ab}^{-1}$ of luminosity.}
\label{tab:2opdim6}
\end{table}

Yet again, we see that the network average reproduces the truth values; and while the cross section in the presence of $O_W$ and $O_{3W}$ matches the SM value by construction, the polarization fraction is clearly different. Plugging in numbers, the $200\,\gev \le p_T \le 300\, \gev$ polarization fraction is different than its SM value by $\mathcal O(5\,\sigma_{f_L})$ using the full luminosity HL-LHC uncertainty, or $\mathcal O(1.5\, \sigma_{f_L})$ using $300\, \text{fb}^{-1}$ values. 

\section{Discussion}
\label{sec:discussion}

In this paper, we have shown how a CNN can be used as a polarization analyzer for hadronic $W^{\pm}$ bosons. The algorithm cannot distinguish between events accurately enough that it can be used as an event-by-event tagger, though this inability to perfectly separate polarizations is not a failure of the network and is present even at parton level. While event-by-event tagging is inefficient, we showed that a template analysis comparing the network average for an unknown sample to the average output of validation samples, does accurately reveal the polarization composition. A benefit of the CNN method is that  it keeps all events without reducing the discriminating power. This can be compared substructure based polarization analyzers, which introduce further cuts on top of the base kinematic selection of boosted $W$\@. In keeping more events while maintaining discriminating power, the uncertainty on the extracted polarization is reduced, and fewer cuts means less concern of reintroducing interference between the different $W$ polarizations.

 We tested the network average method on $pp \to W^{\pm}(jj)Z(\ell\ell)$ production in the SM and in the presence of dimension-6 operators that impact different polarizations. In all cases, we find that the network average reproduces the truth level result and captures how the dimension-6 operator structure dictates how the boson polarizations are affected. Using pseudoexperiments to estimate the uncertainty on $f_L$, we find $\sigma_{f_L}$ at the percent level assuming $3\, \text{ab}^{-1}$ of data, or $\sim 10\%$ for $150\, \text{fb}^{-1}$, with the higher $p_T$ samples having slightly larger uncertainties due to lower statistics.  For the $3\, \text{ab}^{-1}$ estimates, these uncertainties are small compared to the deviations from the SM polarization fraction when the dimension-6 operator $\mathcal O_W$ is included with $c_W = 0.001$, and comparable to the $f_L$ deviations from including $\mathcal O_{3W}, c_{3W} = 0.001$. Furthermore, we find similar uncertainties in scenarios where the $c_W, c_{3W}$ have been chosen to cancel in the total cross section -- a scenario where polarization analysis is the discovery tool for new physics. We obtain these results  results with training sizes of  $\sim \mathcal O$ (few 100K events); the training sample size could be enlarged in future work and may lead to better performance. Finally, the uncertainty estimates above are optimistic, as we have not considered the impact of cuts required to separate reducible SM backgrounds such as $Z(\ell\ell) + \text{jets}$ or pollution from those backgrounds. However, our analysis demonstrates the utility of network-based hadronic $W^\pm$  polarization analyses. 
 
Targets for future study include other processes, such as vector boson scattering, and hadronic $Z$-tagging possibilities. It would also be interesting to explore what information the CNN uses besides the $\cos{\theta^*}$ or $p_\theta$ variable, perhaps by an adversarial network, or to combine $W^\pm$ vs. QCD differentiation and polarization analysis into a single network. 

\section*{Acknowledgments}

We thank Bryan Ostdiek for numerous helpful discussions. We also thank the Center for Research Computing (CRC) at Notre Dame for resources and continuous support.
The work of AM was supported in part by
the National Science Foundation
under Grant Number PHY-1820860.

\appendix
\section{Different Network example}
 \label{appendix}

In addition to the CNN, we tested the performance of two other networks, MaxOut and ResNet. 
\begin{itemize}
\item In a MaxOut network, images are flattened into a vector of inputs, then fed through special `MaxOut' layers that combine nearby inputs in several ways and output the maximum combination \cite{Oliveira_2016}, a process designed to capture some proximity information on neighboring inputs,  to inhibit the sparsity of hidden layer values and assists the dropout layer as shown in~\cite{goodfellow2013maxout}. For the problem at hand, we use a network with two sequential MaxOut layers, the first with 256 units and the second with 128. The second MaxOut layer is followed by 64 and 25 fully connected dense layers with ReLU activation and single output layer with sigmoid function as activation.

\item ResNet networks are image based and are grouped into `residual blocks'. Within each block, the input is processed by several convolution layers, then connected back to the original image. This `bypassing' step was designed to minimize vanishing gradient issues, but comes at the price of increased complexity and thus more trainable parameters. After a number of blocks, the ResNet output is flattened and processed by dense layers.  The ResNet structures we ended up with is shown below in Fig.~\ref{fig:resnet}. 
\end{itemize}

For both networks, we supplemented the architecture with several Dropout layers. These were added, especially for ResNet, to avoid overtraining.

\begin{figure}[h!]
	\centering
	\includegraphics[width=0.4\textwidth]{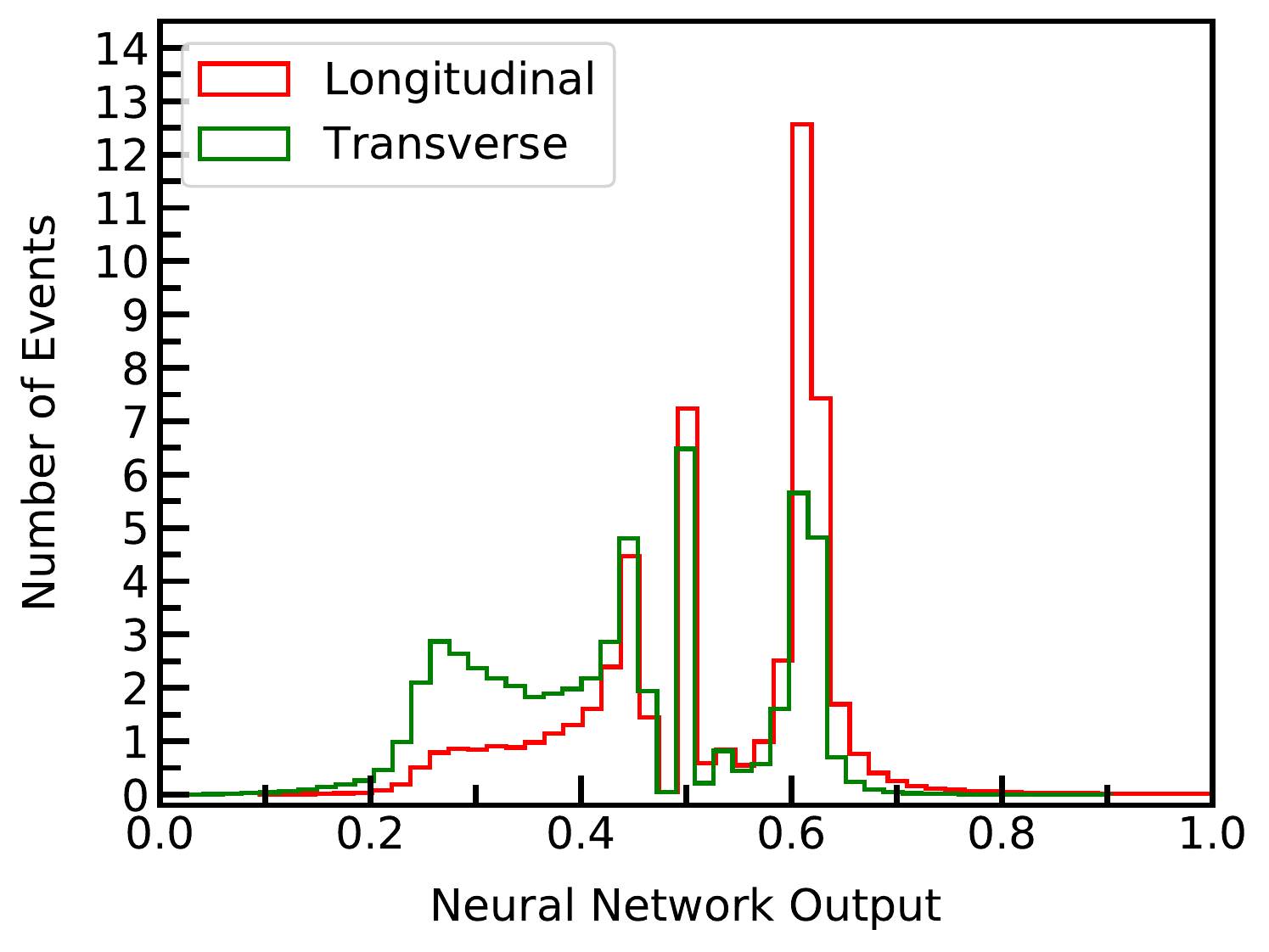}
	\includegraphics[width=0.4\textwidth]{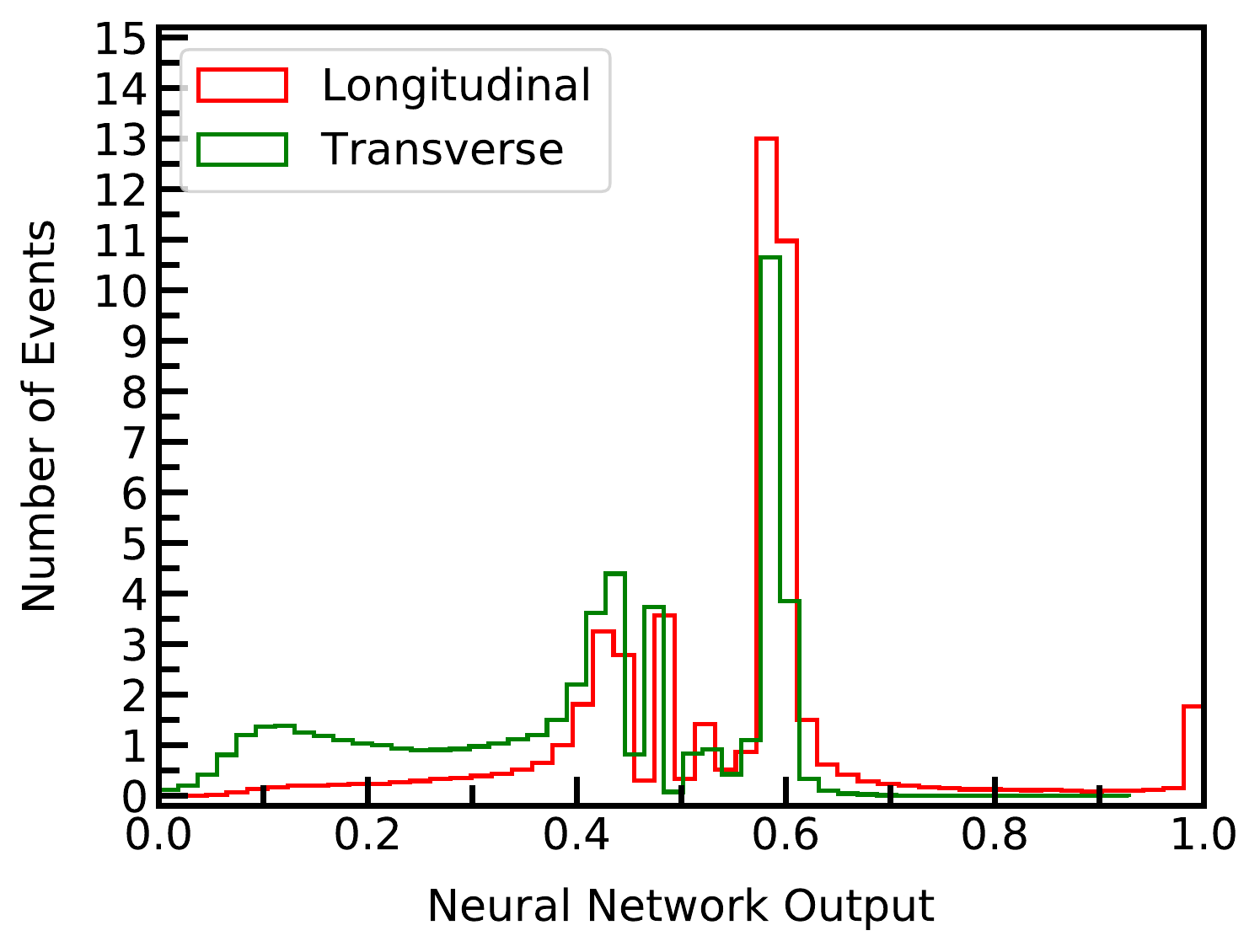}
	\caption{MaxOut distribution result for both $p_T$ bins: left $[200\,\gev, 300\,\gev]$, and right $[400\,\gev, 500\,\gev]$.}
	\label{fig:maxout}
\end{figure}

We train and test our MaxOut and ResNet networks in the same fashion as the CNN discussed in Sec.~\ref{sec:nntrain}. Repeating the $f_L$ and $\delta f_L$ calculations on these comparison networks, we can compare results with the CNN. The extracted $f_L$ values from the CNN, MaxOut and ResNet are presented in Tables \ref{tab:othernet_result}. While the network outputs are different (Fig.~\ref{fig:maxout} shows the MaxOut output for both $p_T$ bins), all three networks perform similarly. These results indicate that the network performance of predicting $f_L$ has reached a saturation point, in the sense that additional network complexity does not yield more accurate results. 

\begin{table}[h!]
	\centering
	
	\begin{tabular}{|c|c|c|c|c|}
		\hline
		 &truth &$f_L$ CNN  &$f_L$ MaxOut   & $f_L$ ResNet      \\ \hline
		SM & 0.265 & $0.259\pm 0.013$ &$0.287 \pm 0.011 $ &   $ 0.259 \pm 0.012 $\\ \hline
		$O_W$ & 0.311& $0.297\pm0.010$ &$ 0.321 \pm 0.010 $   &  $ 0.295 \pm 0.009 $  \\ \hline
		$O_{3W}$ & 0.258 & $0.254\pm0.011$ & $0.282 \pm 0.012$     &  $ 0.257   \pm 0.011  $ \\ \hline
	\end{tabular}	
	\caption{$f_L$ predictions at low $p_T$ $\in$ $[400\,\gev, 500\,\gev]$ for MaxOut, ResNet, and the CNN developed here, along with and the truth value from {\tt MadGraph5v2.7.0}. The errors on $f_L$ have been calculated using the method described in Sec.~\ref{sec:SMWZ} and assuming $3$ ab$^{-1}$ of data.}
	\label{tab:othernet_result_lowpt}
\end{table}

\begin{table}[h!]
	\centering

	\begin{tabular}{|c|c|c|c|c|}
		\hline
		& truth & $f_L$ CNN & $f_L$ MaxOut      &    $f_L$ ResNet       \\ \hline
		SM  & $0.304$ & $0.300 \pm 0.033$ & $0.323 \pm 0.026 $ & $0.301 \pm 0.034 $ \\ \hline
		$O_W$ & $0.439$ & $0.391\pm0.033$ & $0.407 \pm 0.025$ & $0.414 \pm 0.034 $ \\ \hline
		$O_{3W}$ & $0.198$ & $0.181\pm 0.043$ &  $ 0.250 \pm 0.026$ & $0.194 \pm 0.032 $ \\ \hline
	\end{tabular}
	\caption{$f_L$ predictions at high $p_T$ $\in$ $[400\,\gev, 500\,\gev]$ for MaxOut, ResNet, and the CNN developed here, along with and the truth value from {\tt MadGraph5v2.7.0}. The errors on $f_L$ have been calculated using the method described in Sec.~\ref{sec:SMWZ} and assuming $3$ ab$^{-1}$ of data.}
	\label{tab:othernet_result}
\end{table}

\newpage

\begin{figure}[h!]
	\centering
	\includegraphics[width=0.6\textwidth]{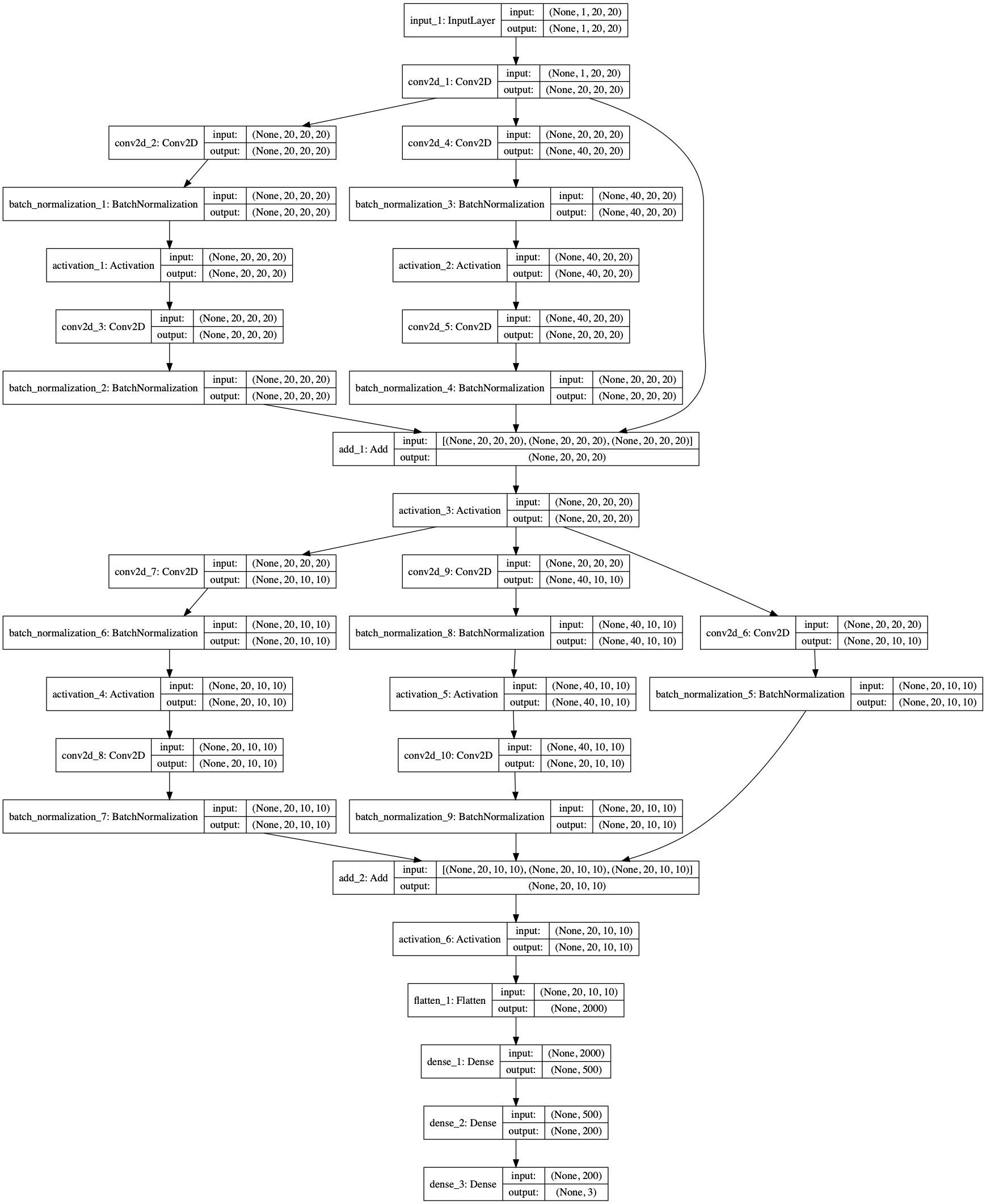}
	\caption{For ResNet Structure, we stack several ResNet blocks with the network shown above. Output of the first block yields the same dimension as the original image and second block deduces the dimension. After the deduction, the convoluted images is followed by flattening and dense network to produce a single output.}
	\label{fig:resnet}
\end{figure}

\bibliographystyle{utphys}
\bibliography{Wpolbib}

\end{document}